# Acoustically modulated optical emission of hexagonal boron nitride layers

Fernando Iikawa[1,2], Alberto Hernández-Mínguez[1,*], Igor Aharonovich[3], Siamak Nakhaie[1], Yi-Ting Liou[1], João Marcelo J. Lopes[1], and Paulo V. Santos[1]

[1]Paul-Drude-Institut für Festkörperelektronik, Leibniz-Institut im Forschungsverbund Berlin e.V., Hausvogteiplatz 5-7, 10117 Berlin, Germany
[2]Institute of Physics, State University of Campinas, 13083-859, Campinas-SP, Brazil
[3]School of Mathematical and Physical Sciences, University of Technology Sydney, Ultimo, New South Wales 2007, Australia

We investigate the effect of surface acoustic waves on the atomic-like optical emission from defect centers in hexagonal boron nitride layers deposited on the surface of a LiNbO$_3$ substrate. The dynamic strain field of the surface acoustic waves modulates the emission lines resulting in intensity variations as large as 50% and oscillations of the emission energy with an amplitude of almost 1 meV. From a systematic study of the dependence of the modulation on the acoustic wave power, we determine a hydrostatic deformation potential for defect centers in this two-dimensional material of about 40 meV/%. Furthermore, we show that the dynamic piezoelectric field of the acoustic wave could contribute to the stabilization of the optical properties of these centers. Our results show that surface acoustic waves are a powerful tool to modulate and control the electronic states of two-dimensional materials.

Defect centers in solids have attracted much attention recently due to their atomic-like optical emission characterized by strong and sharp lines. Typical examples are the nitrogen–vacancy defects in diamond [1, 2] and defect centers in SiC [3]. These centers act as single photon sources thus becoming promising candidates for application in quantum information processing. One of the challenges related to these luminescence centers in solids is to find mechanisms for the efficient control of their optoelectronic properties. To this end, surface acoustic waves (SAWs) are an interesting approach, because their strain and piezoelectric fields oscillating with frequencies in the range of hundreds of MHz can couple efficiently to defect centers placed close to the surface of the vibrating substrate [4, 5, 6].

In this manuscript, we discuss the dynamic modulation of optically active centers in hexagonal boron nitride (h-BN) by SAWs. h-BN is a two-dimensional (2D) crystal with a graphene-like honeycomb atomic lattice. Contrary to graphene, h-BN displays a wide energy band gap (~6 eV), which makes it an exceptional insulator in e.g. van der Waals heterostructures [7, 8]. It has recently been demonstrated that h-BN can host defect centers acting as single photon emitters in both the visible [9, 10, 11, 12, 13, 14, 15, 16, 17, 18] [19, 20, 21] and ultraviolet spectral ranges [22], making this material promising for quantum optics. To this end, it is necessary to develop techniques for the tuning of their emission energies. This can be achieved by applying either static strain [23, 24] or electric fields [25, 26, 27, 28]. Here, we demonstrate that SAWs can couple to defect centers contained in two kinds of h-BN samples, namely multi-layer-thick flakes and few-layer-thick films.

The samples containing multilayer h-BN flakes with a thickness of about 200 nm and a lateral size of about 1 μm (obtained from Graphene Supermarket) were prepared by drop-casting the flakes on a 127°Y-cut LiNbO$_3$ substrate, followed by an annealing step at 850 °C for 30 min. in an

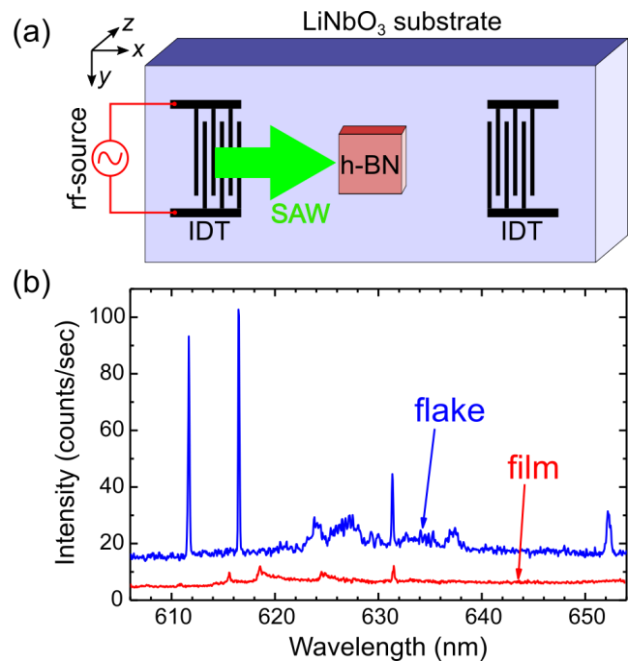

*Fig. 1. (**a**) Schematic diagram of the samples. They contain two interdigital transducers (IDTs) patterned at the surface of LiNbO$_3$. An rf signal applied to one of the IDTs excites a SAW propagating along the region where the h-BN is deposited. (**b**) Low temperature (5 K) luminescence spectrum of defect centers in a multi-layer-thick h-BN flake (blue curve) and in a few-layer-thick film (red curve).*

Argon atmosphere (1 Torr) to activate their emission properties [9, 10]. Finally, SAW delay lines consisting of pairs of interdigital transducers (IDTs) were patterned on the LiNbO$_3$ by optical lithography and lift-off metallization, cf. Fig. 1(a). In contrast, the few-layer-thick h-BN films (about 1 nm thick) were grown on nickel by molecular beam epitaxy (MBE) [29]. Then, an area of about 7×6 mm$^2$ was transferred onto the LiNbO$_3$ using a wet transfer technique [30]. In this case, the LiNbO$_3$ was patterned with acoustic delay lines before transferring the h-BN film. In addition, we did not perform any post-transfer annealing in these samples. This allowed us to directly compare their optical performance with previous results on $SiO_2/Si$, where the MBE-grown films were also not annealed [30].

The samples were investigated in a He cryostat adapted for micro-photoluminescence (μ-PL) measurements with a spatial resolution of about one μm (Attocube Confocal Microscope). The experiments were performed at a nominal temperature of 5 K. The defect centers were optically excited by a 532 nm solid-state laser beam focused onto the sample using an objective with large numerical aperture of about 0.8. The emitted light was collected by the same objective, coupled into a single-mode optical fiber, and sent into a 0.5 m-long monochromator equipped with a 900 mm$^{-1}$ grating and a Si-based charge-coupled device (CCD) camera. The SAWs were excited by applying an rf signal of the appropriate frequency to one of the IDTs using an rf generator connected to an amplifier (about 23 dB amplification).

To detect SAW-induced changes in the PL spectra with the required large signal/noise ratio, we used the modulation method described in Ref. [31]. Here both, rf source and excitation laser, are amplitude-modulated with a modulation frequency of about 300 Hz. The spectra are recorded with SAW and laser excitation modulated *out of phase*, i.e., the sample is exposed to the laser when the rf signal is switched off (SAW-OFF spectrum, $I_{off}$), as well as *in phase* under laser excitation of the sample when the rf signal is switched on (SAW-ON spectrum, $I_{on}$). We repeat accumulation sequences SAW-OFF – SAW-ON – SAW-ON − SAW-OFF over typically 20 times and average the corresponding ON and OFF measurements (see Supplementary Material). This acquisition method increases the signal/noise ratio due to the long accumulation times and, most important, minimizes the effects of systematic temperature and laser power fluctuations, as well as those related to spectral wandering typically observed in h-BN emission centers [15].

The blue curve in Fig. 1(b) shows a typical PL spectrum of a h-BN flake, which consists of several sharp lines distributed along the visible spectral range (between 550 and 800 nm) [11, 12]. We observed similar sharp lines in the transferred h-BN films, cf. red curve in Fig. 1(b), but with a much weaker emission intensity. There are several possible reasons for this weaker emission. First, the flakes were annealed to activate luminescent defects, while the films did not undergo such a process. Second, the flakes contain much more layers than the film. This means that, in principle, the number of defects which are excited by the laser spot and emit light at a certain wavelength should be larger in the flakes than in the film. However, light centers in h-BN show typically a large spectral distribution [11, 15]. Therefore, the probability of finding two or more centers emitting at exactly the same wavelength is relatively low. Finally, the interaction of the defect centers with the $LiNbO_3$ substrate may also play a role. Actually, the intensity of the emission lines in the MBE-film was weaker than the ones measured under comparable experimental conditions in similar samples transferred to $SiO_2/Si$ [30]. It has been reported that electric-field-induced charging effects can tune the brightness of luminescent centers [25, 32, 33]. As $LiNbO_3$ is a ferroelectric material, its spontaneous polarization fields could weaken the luminescence intensity of h-BN defects close to its surface. This effect is expected to affect the thin films more significantly than the flakes since the thicker flakes contain also defects in larger distance from the $LiNbO_3$, which are therefore optically brighter.

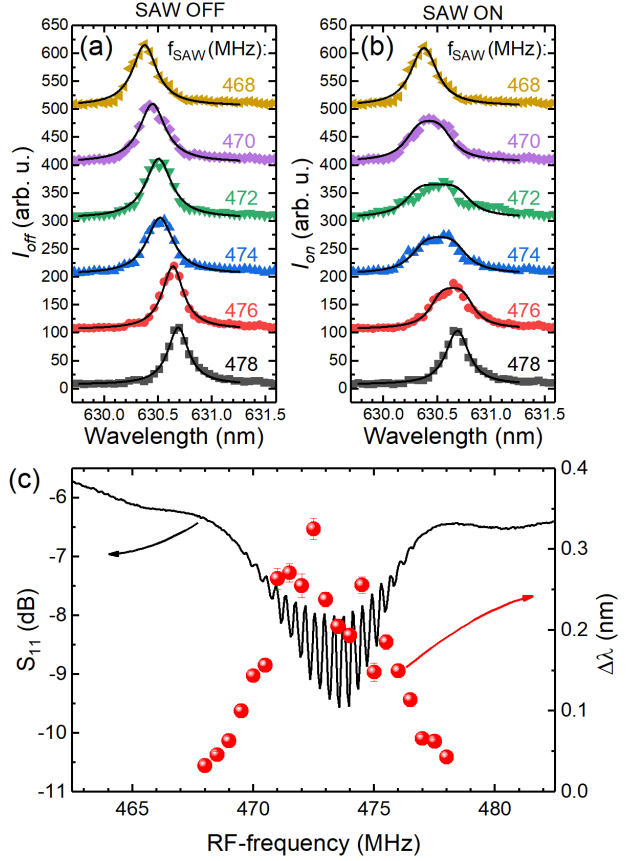

*Fig. 2. (a) $I_{off}$ and (b) $I_{on}$ spectra of a luminescent center as a function of the rf frequency applied to the IDT that launches the SAW, $f_{SAW}$ (nominal rf power of -1 dBm). The black curves are fittings to the data according to Eqs. 1 and 2. The data are vertically shifted for clarity. (c) rf-power reflection coefficient $S_{11}$ of the IDT (black curve), and amplitude of the spectral oscillation, $\Delta\lambda$ (red dots), calculated for several rf frequencies across the resonance of the IDT.*

Figure 2 displays the PL spectra of one of the emission lines of an h-BN flake, recorded in the absence (Fig. 2(a), SAW OFF) and in the presence of SAW excitation (Fig. 2(b), SAW ON). The measurements were performed for several frequencies of the rf signal applied to the IDT. As the $I_{off}$ spectra are recorded with the exciting laser and rf signal out of phase, we do not expect any effect of the SAW on the emission properties. We have fitted the $I_{off}$ spectra by the following Lorentzian function [black curves in Fig. 2(a)]:

$$I_{off}(\lambda) = I_0 + \frac{2A}{\pi} \frac{w}{4(\lambda-\lambda_c)^2+w^2}, \quad (1)$$

where $A$, $w$ and $\lambda_c$ are the area, width and center of the emission line, respectively, and $I_0$ accounts for the background signal. While the width remains quite stable at an average value of $0.28\pm0.03$ nm, the peak center shifts from 630.4 nm to 630.7 nm during the experiment. Spectra measured as a function of time (see Supplementary Material) indicate that this shift is a consequence of spectral and intensity fluctuations (spectral diffusion) probably caused by aleatory changes in the ionization state of nearby charge traps like shallow impurities or defects [15, 34, 35, 36, 37]. These fluctuations are also responsible for the relatively broad emission line and for the partial deviation of the line shape from that of a single peak.

The corresponding $I_{on}$ spectra, in contrast, show an additional line broadening for a certain range of rf frequencies, accompanied by a decrease in the amplitude. We interpret this change in the shape of the line as caused by the dynamic strain field of the SAW: the oscillating tensile and compressive strain modulates periodically the crystal structure of the luminescent center, thus leading to the oscillation of its optical transition energy around the equilibrium value [23, 24]. As the acquisition time of the µ-PL spectra is much larger than the SAW period, the energy oscillation manifests as a broadening of the emission line and a reduction of its maximum intensity, as shown in Fig. 2(b). Similar broadening effects under dynamic strain have been commonly observed in the luminescence spectra of quantum dots [38, 39, 40, 41], as well as in the electronic spin resonance of defects in diamond [42], and in the Raman lines of semiconductors [31].

To quantify the strength of the optomechanical coupling, we have fitted the $I_{on}$ spectra of Fig. 2(b) according to the formula [39, 42, 43, 44]:

$$I_{on}(\lambda) = I_0 + \\ + f_{SAW} \int_0^{1/f_{SAW}} \frac{2A}{\pi} \frac{w}{4\{\lambda-[\lambda_c+\Delta\lambda\cos(2\pi f_{SAW}t)]\}^2+w^2} dt. \quad (2)$$

Here, the term in the integrand is the Lorentzian function with the values for $A$, $w$ and $\lambda_c$ obtained from the fitting of the corresponding $I_{off}$ spectra in Fig. 2(a), and $f_{SAW}$ is the frequency of the rf signal that launches the SAW. The amplitude of the SAW-induced spectral oscillation is given by $\Delta\lambda$, which is the only free parameter in the fitting. We display in Fig. 2(c) the dependence of $\Delta\lambda$ on the rf frequency (red dots), and compare it with the rf-power reflection coefficient, $S_{11}$, of the IDT (black curve, measured using a vector network analyzer). Non-vanishing values of $\Delta\lambda$ are only observed for rf frequencies within the minimum in the $S_{11}$ spectrum, corresponding to the frequency range for efficient SAW excitation by the IDT. The maximum intensity suppression (of about 50%) and broadening of the PL emission occur for SAW frequencies around 473 MHz, and are well reproduced by our fitting supposing a value of $\Delta\lambda$ up to 0.3 nm. This corresponds to SAW-induced oscillations of the emission energy with amplitudes of almost 1 meV.

As $LiNbO_3$ is a strong piezoelectric material, the SAW strain field is accompanied by an oscillating electric field that could also, in principle, be responsible for the modulation of the emission line. It has recently been reported that electric fields of 1 GV/m can induce Stark shifts in h-BN defect centers on the order of 10 nm [25, 26, 28]. However, the SAW-induced piezoelectric fields in $LiNbO_3$ are typically on the order of 1 MV/m, which means that the expected Stark shifts are about 0.01 nm. Therefore, we can neglect the contribution of the SAW piezoelectric fields to the observed modulation.

Another evidence supporting the preeminent role of the strain over the piezoelectric field in the acoustic modulation is the fact that, out of twenty studied emission centers, only three effectively underwent the acoustic modulation. This indicates that the quality of the physical contact between the flakes and the $LiNbO_3$ substrate is critical for the coupling of the SAW into the h-BN. In addition, the amount of strain that can be transmitted from the bottom to the top of the multilayer flakes can be strongly limited due to the weak Van-der-Waals interaction between adjacent monolayers in 2D materials [23, 45]. As a matter of fact, the emission centers undergoing acoustic modulation were optically weaker than the unperturbed ones, thus supporting our assumption that the LiNbO$_3$ substrate affects the optical properties of emission centers in close contact to the substrate.

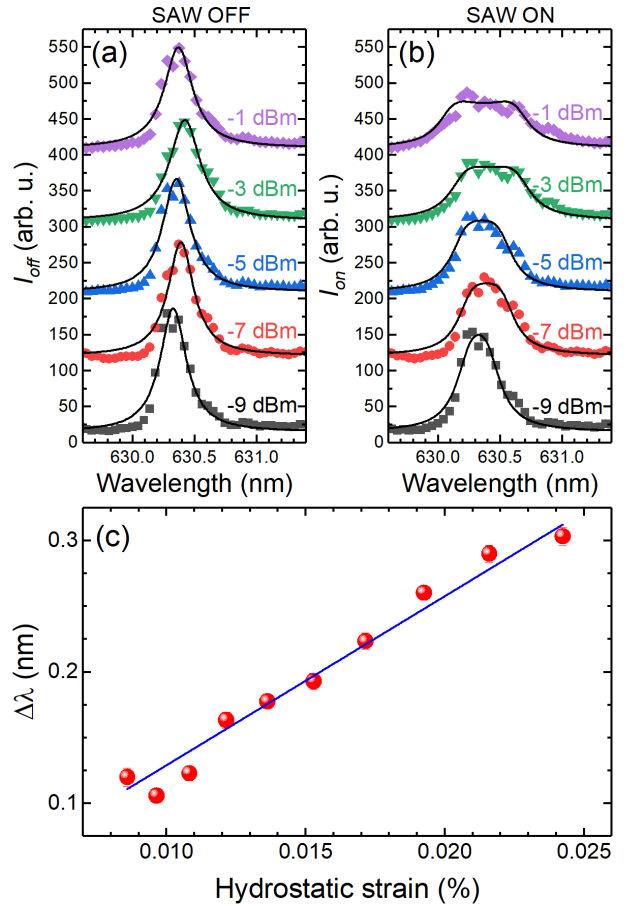

*Fig. 3. (a) $I_{off}$ and (b) $I_{on}$ spectra as a function of nominal rf-power (recorded for $f_{SAW}$ = 473.3 MHz). The black curves are fits according to Eqs. 1 and 2. The data are vertically shifted for clarity. (c) Amplitude of the spectral oscillation, $\Delta\lambda$, with respect to the amplitude of the SAW-induced hydrostatic strain at the surface of the $LiNbO_3$. The blue line is a linear fit to the data.*

Although all three centers undergoing acoustic modulation showed similar behavior, only the one discussed in Fig. 2 was stable enough to perform a systematic study of the SAW-induced effects on the emission line. Figures 3(a) and 3(b) show the dependence of the $I_{off}$ and $I_{on}$ spectra on the nominal rf power applied by the rf generator. As expected, the $I_{off}$ spectra remain unperturbed by the SAW, while the SAW-induced broadening of the $I_{on}$ spectra increases with the amplitude of the rf power. Figure 3(c) displays the fitted values of $\Delta\lambda$ as a function of the amplitude of the hydrostatic strain, $\varepsilon_0 = \varepsilon_{xx} + \varepsilon_{zz}$, at the surface of the LiNbO$_3$ for the different nominal rf powers (since the SAW is a Rayleigh mode, strain is generated along the $x$ and $z$ directions, but not along the $y$ direction [46]). To determine the values for $\varepsilon_0$, we first calculated the SAW power density from the nominal rf power applied to the IDT and the reflection coefficient of Fig. 2(c). We then related the SAW power density to $\varepsilon_0$ by solving numerically the coupled elastic and electromagnetic equations for the LiNbO$_3$ substrate. As expected, $\Delta\lambda$ increases linearly with $\varepsilon_0$ at a rate of 12.9±0.2 nm per % of strain, which corresponds to a deformation potential of about 40 meV/%. This value is ten times larger than the ones reported for multilayer h-BN under static strain [23, 24], but similar to the result recently obtained for SAW-modulated defects in h-BN powder [47].

Finally, we discuss the MBE-grown h-BN films. As mentioned before, the emission lines of these samples are surprisingly weak. In addition, they undergo significant spectral diffusion. These features masked the observation of the acoustic broadening reported above for the flakes. Under the application of the SAW, however, we observed a stabilization of the optical emission for some light centers. As an example, Fig. 4(a) shows a series of PL spectra recorded as a function of time (time delay of 20 s between two successive acquisitions) for a sharp peak emitting at 624 nm in the absence of SAWs. The emission wavelength fluctuates in time between two well-defined values at 623.5 and 624.1 nm. The intensity of the emission line also fluctuates, even though laser excitation density and sample temperature were kept constant during the experiment. Figure 4(b) shows the same sequence of spectra in the presence of a SAW of $f_{SAW}$=513.50 MHz and nominal rf power of 0 dBm (the IDTs in this sample were designed to launch SAWs with a different frequency than the ones for the flakes). Under these experimental conditions, the 623.5 nm peak disappeared. This effect was systematically observed in three out of ten luminescent centers studied, and it was reproducible every time that we switched on the acoustic waves.

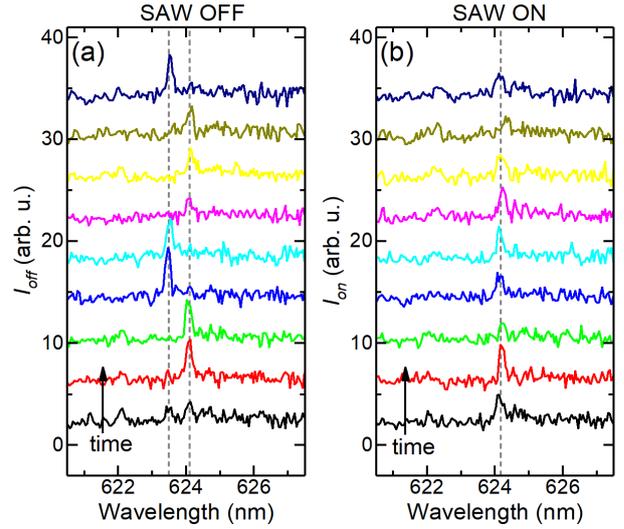

*Fig. 4. Time evolution of the light emitted by a center in the MBE-grown film (**a**) in the absence of SAW, and (**b**) when the SAW is applied. The data are vertically shifted for clarity. The spectra were recorded sequentially starting from the bottom (integration time of 20 s per spectrum). The vertical dashed lines mark the average positions of the peaks.*

Following the model discussed before for the flakes, the spectral fluctuations in absence of SAWs suggest a strong coupling of the luminescent center to the ionization state of a nearby shallow charge trap [37, 48]. Therefore, we interpret the stabilization of the luminescence as caused by the interaction of the SAW with the nearby charge trap. It is known that the piezoelectric fields of SAWs can control the charge population of shallow quantum-dot-like centers by dynamically injecting and/or extracting charge carriers [49, 50, 51, 52, 41]. According to this, we attribute the quenching of the 623.5 nm peak in Fig. 4(b) to the fact that the dynamic SAW piezoelectric field favors a certain ionization state of the nearby shallow trap. This can happen either by continuously injecting charge carriers into the trap, or by extracting the charge carriers as soon as they are trapped. Supposing that the high and low energy peaks in Fig. 4(a) reflect the interaction of the light center with a nearby shallow trap in its negative and neutral charge states, respectively, then the suppression of the high energy peak in Fig. 4(b) would indicate that charge carrier extraction is the dominant mechanism in this case. Although a more comprehensive understanding of this mechanism will require additional studies that go beyond the scope of this manuscript, this result suggests that SAWs could be a helpful tool to reduce the energy fluctuation of the emission lines in the case of unstable emitters, as in our MBE-grown h-BN film.

In conclusion, we have investigated the interaction of SAWs with optically active defect centers in h-BN flakes and films transferred to the surface of LiNbO$_3$. In the case of the flakes, we have demonstrated the modulation of the emission lines by the alternating acoustic field and estimated a deformation potential for the defects of about 40 meV/%. For the MBE-grown film, although the weakness of the emission lines and their strong spectral diffusion prevented us to observe the acoustically induced broadening, the presence of the SAW fields suppressed the spectral fluctuations, thus leading to a more stable optical emission of the centers.

**Supplementary Material**

See Supplementary Material for luminescence spectra as a function of time for a light center in h-BN flake.

**Acknowledgements**

The authors would like to thank Dr. Snežana Lazić for discussions and Dr. Lutz Schrottke for a critical reading of the manuscript. F.I. and I.A. acknowledge Alexander von Humboldt Foundation for financial support. F.I. acknowledges Conselho Nacional de Desenvolvimento Científico e Tecnológico (305769/2015-4 and 432882/2018-9) for financial support.

*alberto.h.minguez@pdi-berlin.de

**References**

[1] C. Kurtsiefer, S. Mayer, P. Zarda and H. Weinfurter, "Stable Solid-State Source of Single Photons," *Phys. Rev. Lett.,* vol. 85, pp. 290-293, 2000.

[2] L. Childress and R. Hanson, "Diamond NV centers for quantum computing and quantum networks," *MRS Bulletin,* vol. 38, p. 134, 2013.

[3] W. F. Koehl, B. B. Buckley, F. J. Heremans, G. Calusine and D. A. David, "Room temperature coherent control of defect spin qubits in silicon carbide," *Nature,* vol. 479, pp. 84-87, 2011.

[4] D. A. Golter, T. Oo, M. Amezcua, K. A. Stewart and H. Wang, "Optomechanical Quantum Control of a Nitrogen-Vacancy Center in Diamond," *Phys. Rev. Lett.,* vol. 116, p. 143602, 2016.

[5] D. A. Golter, T. Oo, M. Amezcua, I. Lekavicius, K. A. Stewart and H. Wang, "Coupling a Surface Acoustic Wave to an Electron Spin in Diamond via a Dark State," *Phys. Rev. X,* vol. 6, p. 041060, 2016.

[6] S. J. Whiteley, G. Wolfowicz, C. P. Anderson, A. Bourassa, H. Ma, M. Ye, G. Koolstra, K. J. Satzinger, M. V. Holt, F. J. Heremans, A. N. Cleland, D. I. Schuster, G. Galli and D. D. Awschalom, "Spin-phonon interactions in silicon carbide addressed by Gaussian acoustics," *Nat. Phot.,* 2019.

[7] A. K. Geim and I. V. Grigorieva, "Van der Waals heterostructures," *Nature,* vol. 499, p. 419, 2013.

[8] K. S. Novoselov, A. Mishchenko, A. Carvalho and A. H. Castro Neto, "2D materials and van der Waals heterostructures," *Science,* vol. 353, p. aac9439, 2016.

[9] T. T. Tran, K. Bray, M. J. Ford, M. Toth and I. Aharonovich, "Quantum emission from hexagonal boron nitride monolayers," *Nat. Nanotechnol.,* vol. 11, p. 37, 2016.

[10] T. T. Tran, C. Zachreson, A. M. Berhane, K. Bray, R. G. Sandstrom, L. H. Li, T. Taniguchi, K. Watanabe, I. Aharonovich and M. Toth, "Quantum Emission from Defects in Single-Crystalline Hexagonal Boron Nitride," *Phys. Rev. Appl.,* vol. 5, p. 034005, 2016.

[11] T. T. Tran, C. Elbadawi, D. Totonjian, C. J. Lobo, G. Grosso, H. Moon, D. R. Englund, M. J. Ford, I. Aharonovich and M. Toth, "Robust Multicolor Single Photon Emission from Point Defects in Hexagonal Boron Nitride," *ACS Nano,* vol. 10, p. 7331, 2016.

[12] N. R. Jungwirth, B. Calderon, Y. Ji, M. G. Spencer, M. E. Flatté and G. D. Fuchs, "Temperature Dependence of Wavelength Selectable Zero-Phonon Emission from Single Defects in Hexagonal Boron Nitride," *Nano Lett.,* vol. 16, p. 6052, 2016.

[13] L. J. Martínez, T. Pelini, V. Waselowski, J. R. Maze, B. Gil, G. Cassabois and V. Jacques, "Efficient single photon emission from a high-purity hexagonal boron nitride crystal," *Phys. Rev. B,* vol. 94, p. 121405(R), 2016.

[14] N. Chejanovsky, M. Rezai, F. Paolucci, Y. Kim, T. Rendler, W. Rouabeh, F. Fávaro de Oliveira, P. Herlinger, A. Denisenko, S. Yang, I. Gerhardt, A. Finkler, J. H. Smet and J. Wratchup, "Structural Attributes and Photodynamics of Visible Spectrum Quantum Emitters in Hexagonal Boron Nitride," *Nano Lett.,* vol. 16, p. 7037, 2016.

[15] Z. Shotan, H. Jayakumar, C. R. Considine, M. Mackoit, H. Fedder, J. Wratchup, A. Alkauskas, M. W. Doherty, V. M. Menon and C. A. Meriles, "Photoinduced Modification of Single-Photon

Emitters in Hexagonal Boron Nitride," *ACS Photonics,* vol. 3, p. 2490, 2016.

[16] B. Sontheimer, M. Braun, N. Nikolay, N. Sadzak, I. Aharonovich and O. Benson, "Photodynamics of quantum emitters in hexagonal boron nitride revealed by low-temperature spectroscopy," *Phys. Rev. B,* vol. 96, p. 121202(R), 2017.

[17] A. L. Exarhos, D. A. Hopper, R. R. Grote, A. Alkauskas and L. C. Bassett, "Optical Signatures of Quantum Emitters in Suspended Hexagonal Boron Nitride," *ACS Nano,* vol. 11, p. 3328, 2017.

[18] M. Kianinia, S. A. Tawfik, B. Regan, T. T. Tran, M. J. Ford, I. Aharonovich and M. Toth, "Robust Solid State Quantum System Operating at 800 K," *ACS Phot.,* vol. 4, p. 768, 2017.

[19] N. V. Proscia, Z. Shotan, H. Jayakumar, P. Reddy, C. Cohen, M. Dollar, A. Alkauskas, M. Doherty, C. A. Meriles and V. M. Menon, "Near-deterministic activation of room-temperature quantum emitters in hexagonal boron nitride," *Optica,* vol. 5, p. 1128, 2018.

[20] A. L. Exarhos, D. A. Hopper, R. N. Patel, M. W. Doherty and L. C. Bassett, "Magnetic-field-dependent quantum emission in hexagonal boron nitride at room temperature," *Nat. Comm.,* vol. 10, p. 222, 2019.

[21] M. Koperski, K. Nogajewski and M. Potemski, "Single photon emitters in boron nitride: More than a supplementary material," *Optics Comm.,* vol. 411, p. 158, 2018.

[22] R. Bourrellier, S. Meuret, A. Tararan, O. Stéphan, M. Kociak, L. H. G. Tizei and A. Zobelli, "Bright UV Single Photon Emission at Point Defects in h-BN," *Nano Lett.,* vol. 16, p. 4317, 2016.

[23] G. Grosso, H. Moon, B. Lienhard, S. Ali, D. K. Efetov, M. M. Furchi, P. Jarillo-Herrero, M. J. Ford, I. Aharonovich and D. Englund, "Tunable and high-purity room temperature single-photon emission from atomic defects in hexagonal boron nitride," *Nat. Comm.,* vol. 8, p. 705, 2017.

[24] Y. Xue, H. Wang, Q. Tan, J. Zhang, T. Yu, K. Ding, D. Jiang, X. Dou, J. Shi and B. Sun, "Anomalous Pressure Characteristics of Defects in Hexagonal Boron Nitride Flakes," *ACS Nano,* vol. 12, p. 7127, 2018.

[25] G. Noh, D. Choi, J.-H. Kim, D.-G. Im, Y.-H. Kim, H. Seo and J. Lee, "Stark Tuning of Single-Photon Emitters in Hexagonal Boron Nitride," *Nano Lett.,* vol. 18, p. 4710, 2018.

[26] A. Scavuzzo, S. Mangel, J.-H. Park, S. Lee, D. L. Duong, C. Strelow, A. Mews, M. Burghard and K. Kern, "Electrically tunable quantum emitters in an ultrathin graphene-hexagonal boron nitride van der Waals heterostructure," *Appl. Phys. Lett.,* vol. 114, p. 062104, 2019.

[27] N. Mendelson, Z.-Q. Xu, T. T. Tran, M. Kianinia, J. Scott, C. Bradac, I. Aharonovich and M. Toth, "Engineering and Tuning of Quanum Emitters in Few-Layer Hexagonal Boron Nitride," *ACS Nano,* 2019.

[28] N. Nikolay, N. Mendelson, N. Sadzak, F. Böhm, T. T. Tran, B. Sontheimer, I. Aharonovich and O. Benson, "Very Large and Reversible Stark Shift Tuning of Single Emitters in Layered Hexagonal Boron Nitride," *arXiv:1812.02530,* 2019.

[29] S. Nakhaie, M. Heilmann, T. Krause, M. Hanke and J. M. J. Lopes, "Nucleation and growth of atomically thin hexagonal boron nitride on Ni/MgO(111) by molecular beam epitaxy".

[30] A. Hernández-Mínguez, J. Lähnemann, S. Nakhaie, J. M. J. Lopes and P. V. Santos, "Luminescent Defects in a Few-Layer h-BN Film Grown by Molecular Beam Epitaxy," *Phys. Rev. App.,* vol. 10, p. 044031, 2018.

[31] F. Iikawa, A. Hernández-Mínguez, M. Ramsteiner and P. V. Santos, "Optical phonon modulation in semiconductors by surface acoustic waves," *Phys. Rev. B,* vol. 93, p. 195212, 2016.

[32] C. Chakraborty, L. Kinnischtzke, K. M. Goodfellow, R. Beams and A. N. Vamivakas, "Voltage-controlled quantum light from and atomically thin semiconductor," *Nat. Nanotech.,* vol. 10, p. 507, 2015.

[33] C. F. de las Casas, D. J. Christle, J. U. Hassan, T. Ohshima, N. T. Son and D. A. David, "Stark tuning and electrical charge state control of single divacancies in silicon carbide," *Appl. Phys. Lett.,* vol. 111, p. 262403, 2017.

[34] X. Li, G. D. Shepard, A. Cupo, N. Camporeale, K. Shayan, Y. Luo, V. Meunier and S. Strauf, "Nonmagnetic Quantum Emitters in Boron Nitride with Ultranarrow and Sideband-Free Emission Spectra," *ACS Nano,* vol. 11, p. 6652, 2017.

[35] W. P. Ambrose and W. E. Moerner, "Fluorescence spectroscopy and spectral diffusion of single impurity molecules in a crystal," *Nature,* vol. 349, p. 225, 1991.

[36] O. Neitzke, A. Morfa, J. Wolters, A. W. Schell, G. Kewes and O. Benson, "Investigation of Line Width Narrowing and Spectral Jumps of Single Stable Defect Centers in ZnO at Cryogenic Temperature," *Nano Lett.,* vol. 15, p. 3024, 2015.

[37] J. Houel, A. V. Kuhlmann, L. Greuter, F. Xue, M. Poggio, B. D. Gerardot, P. A. Dalgarno, A. Badolato, P. M. Petroff, A. Ludwig, D. Reuter, A. D. Wieck and R. J. Warburton, "Probing Single-Charge Fluctuations at a GaAs/AlAs Interface Using Laser Spectroscopy on a Nearby InGaAs Quantum Dot," *Phys. Rev. Lett.,* vol. 108, p. 107401, 2012.

[38] J. R. Gell, M. B. Ward, R. J. Young, R. M. Stevenson, P. Atkinson, D. Anderson, G. A. C. Jones, D. A. Ritchie and A. J. Shields, "Modulation of single quantum dot energy levels by a surface-acoustic-wave," *Appl. Phys. Lett.,* vol. 93, p. 081115, 2008.

[39] I. Yeo, P.-L. de Assis, A. Gloppe, E. Dupont-Ferrier, P. Verlot, N. S. Malik, E. Dupuy, J. Claudon, J.-M. Gérard, A. Auffèves, G. Nogues, S. Seidelin, J.-P. Poizat, O. Arcizet and M. Richard, "Strain-mediated coupling in a quantum dot-mechanical oscillator hybrid system," *Nat. Nanotech.,* vol. 9, p. 106, 2014.

[40] J. Pustiowski, K. Müller, M. Bichler, G. Koblmüller, J. J. Finley, A. Wixforth and H. J. Krenner, "Independent dynamic acousto-mechanical and electrostatic control of individual quantum dots in a LiNbO3-GaAs hybrid," *Appl. Phys. Lett.,* vol. 106, p. 013107, 2015.

[41] S. Lazic, A. Hernández-Mínguez and P. V. Santos, "Control of single photon emitters in semiconductor nanowires by surface acoustic waves," *Semicond. Sci. Technol.,* vol. 32, p. 084002, 2017.

[42] O. Arcizet, V. Jacques, A. Siria, P. Poncharal, P. Vincent and S. Seidelin, "A single NV defect coupled to a nanomechanical oscillator," *Nat. Phys.,* vol. 7, p. 879, 2011.

[43] E. D. S. Nysten, Y. H. Huo, H. Yu, G. F. Song, A. Rastelli and H. J. Krenner, "Multi-harmonic quantum dot optomechanics in fused LiNbO3-(Al)GaAs hybrids," *J. Phys. D: Appl. Phys.,* vol. 50, p. 43LT01, 2017.

[44] S. Lazic, E. Chernysheva, A. Hernández-Mínguez, P. V. Santos and H. P. van der Meulen, "Acoustically regulated optical emission dynamics from quantum dot-like emission centers in GaN/InGaN nanowire heterostructures," *J. Phys. D: Appl. Phys.,* vol. 51, p. 104001, 2018.

[45] H. Kumar, L. Dong and V. B. Shenoy, "Limits of Coherency and Strain Transfer in Flexible 2D van der Waals Heterostructures: Formation of Strain Solitons and Interlayer Debonding," *Sci. Rep.,* vol. 6, p. 21516, 2016.

[46] L. Rayleigh, "On Waves Propagated along the Plane Surface of an Elastic Solid," *Proc. London Math. Soc.,* Vols. s1-17, pp. 4-11, 1885.

[47] S. Lazic, *Internal communication (2018).*

[48] T. Müller, I. Aharonovich, L. Lombez, Y. Alaverdyan, A. N. Vamivakas, S. Castelletto, F. Jelezko, J. Wrachtrup, S. Prawer and M. Atatüre, "Wide-range electrical tunability of single-photon emission from chromium-based colour centres in diamond," *New J. Phys.,* vol. 13, p. 075001, 2011.

[49] S. Völk, F. J. Schülein, F. Knall, D. Reuter, A. D. Wieck, T. A. Truong, H. Kim, P. M. Petroff, A. Wixforth and H. J. Krenner, "Enhanced Sequential Carrier Capture into Individual Quantum Dots and Quantum Posts Controlled by Surface Acoustic Waves," *Nano Lett.,* vol. 10, p. 3399, 2010.

[50] F. J. R. Schülein, K. Müller, M. Bichler, G. Koblmüller, J. J. Finley, A. Wixforth and H. J. Krenner, "Acoustically regulated carrier injection into a single optically active quantum dot," *Phys. Rev. B,* vol. 88, p. 085307, 2013.

[51] M. Weiß, J. B. Kinzel, F. J. R. Schülein, M. Heigl, D. Rudolph, S. Morkötter, M. Döblinger, M. Bichler, G. Abstreiter, J. J. Finley, G. Koblmüller, A. Wixforth and H. J. Krenner.

[52] A. Hernández-Mínguez, M. Möller, S. Breuer, C. Pfüller, C. Somaschini, S. Lazic, O. Brandt, A. García-Cristóbal, M. M. de Lima, Jr., A. Cantarero, L. Geelhaar, H. Riechert and P. V. Santos, "Acoustically Driven Photon Antibunching in Nanowires," *Nano Lett.,* vol. 12, p. 252, 2012.

[53] T. Roy, M. Tosun, Sachid, A. B. Sachid, S. B. Desai, M. Hettick, C. C. Hu and A. Javei, "Field-Effect Transistors Built from All Two-Dimensional Material Components," *ACS Nano,* vol. 8, pp. 6259-6264, 2014.

[54] S. Nakhaie, J. M. Wofford, T. Schumann, U. Jahn, M. Ramsteiner, M. Hanke, J. M. J. Lopes and H. Riechert, "Synthesis of atomically thin hexagonal boron nitride films on nickel foils by molecular beam epitaxy," *Appl. Phys. Lett.,* vol. 106, p. 213108, 2015.

[55] B. N. J. Persson and N. D. Lang, "Electron-hole-pair quenching of excited states near a metal," *Phys. Rev. B,* vol. 26, p. 5409, 1982.

[56] G. W. Ford and W. H. Weber, "Electromagnetic interactions of molecules with metal surfaces," *Phys. Rep.,* vol. 113, p. 195, 1984.

# Supplementary Material
# Acoustically modulated optical emission of hexagonal boron nitride layers

Fernando Iikawa[1,2], Alberto Hernández-Mínguez[1,*], Igor Aharonovich[3], Siamak Nakhaie[1], Yi-Ting Liou[1], João Marcelo J. Lopes[1], and Paulo V. Santos[1]

[1]*Paul-Drude-Institut für Festkörperelektronik, Leibniz-Institut im Forschungsverbund Berlin e.V., Hausvogteiplatz 5-7, 10117 Berlin, Germany*
[2]*Institute of Physics, State University of Campinas, 13083-859, Campinas-SP, Brazil*
[3]*School of Mathematical and Physical Sciences, University of Technology Sydney, Ultimo, New South Wales 2007, Australia*

This Supplementary Material contains experimental data showing the effect of spectral diffusion on the luminescence spectra of the h-BN flakes. Figure S1 shows luminescence spectra as a function of time, measured under SAWs of 473.3 MHz frequency and two different nominal rf powers. The spectra were recorded in three consecutive SAW-OFF – SAW-ON – SAW-ON – SAW-OFF acquisitions series with an accumulation time of 20 s for each acquisition. The data presented in the manuscript are the result of averaging the corresponding SAW-ON (thick curves) and SAW-OFF (thin curves) measurements over twenty OFF–ON–ON–OFF cycles. The light emission fluctuates with time between four spectral positions marked by the vertical dashed lines.

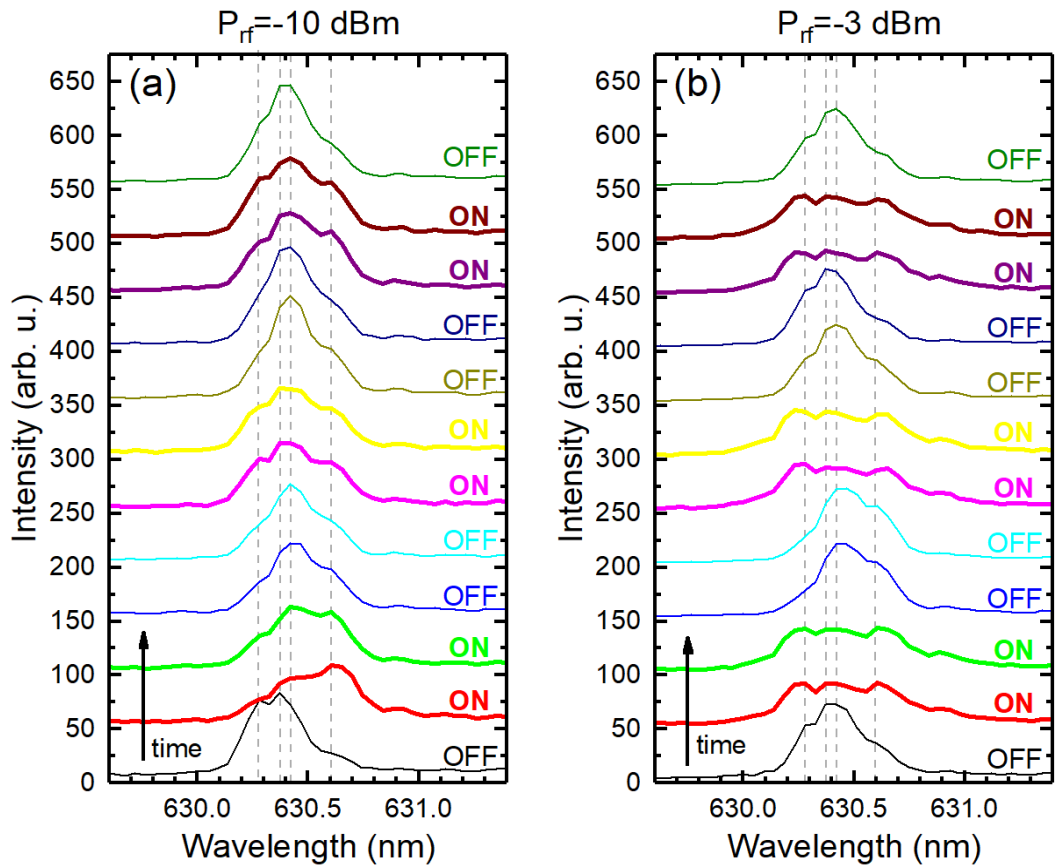

Fig. S1. Time evolution of the light emitted by a luminescence center in an h-BN flake. The spectra were recorded sequentially starting from the bottom. The thin curves show data recorded during the SAW-OFF phase, while the thick curves correspond to the SAW-ON configuration. The emission wavelength center fluctuates with time among the four positions marked as vertical dashed lines.

Fig. S2 shows averaged spectra as measured in Fig. S1, as a function of the rf frequency applied to the interdigital transducer that launches the SAW (nominal rf power of -1 dBm). Due to spectral diffusion, the center of the emission line fluctuates around an average value of 630.4 MHz. By means of the SAW-OFF – SAW-ON – SAW-ON – SAW-OFF acquisition technique, the effect of spectral diffusion is the same in both SAW-ON and SAW-OFF measurements, thus allowing for the identification of the SAW-induced modulation of the emission line as an additional broadening in the region between 470 MHz and 476 MHz.

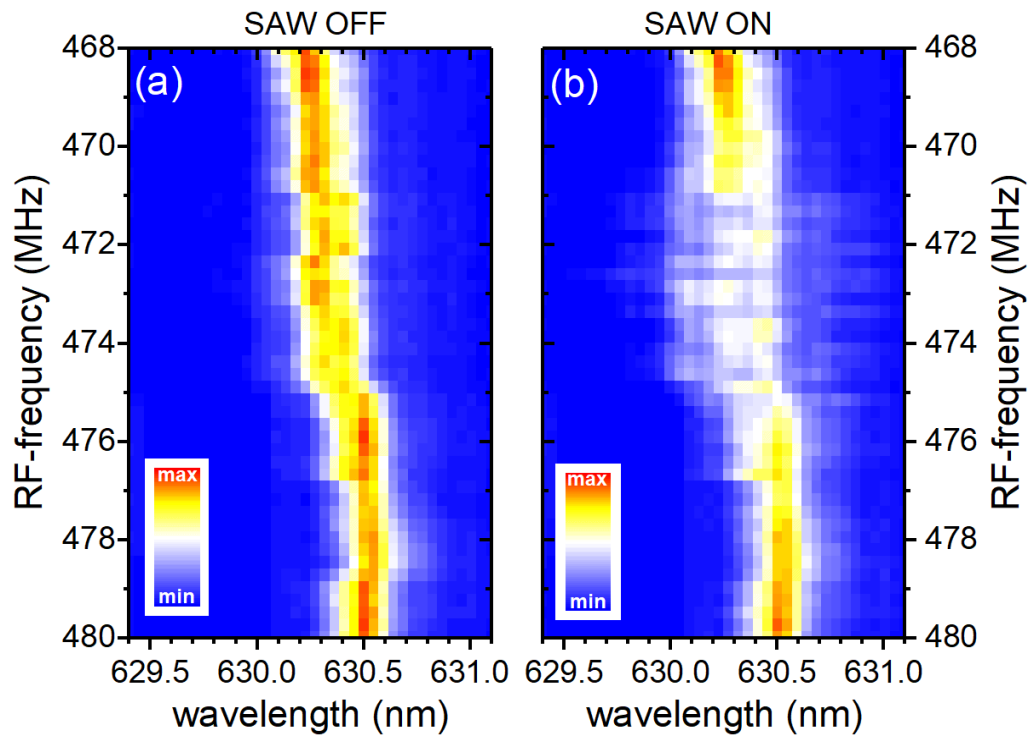

Fig. S2. Luminescence spectra (a) in the absence of SAWs, and (b) in the presence of SAWs, measured for a series of rf frequencies applied to the IDT that launches the SAW. In addition to the effect of spectral diffusion on the wavelength center of the emission line, an additional line broadening for SAW ON in the region between 470 MHz and 476 MHz reveals the coupling of the SAW to the luminescence center.